\documentclass[aps,pra,twocolumn,superscriptaddress]{revtex4-2}
\usepackage{graphicx,color,amsmath,amssymb}

\usepackage{xcolor}
\usepackage[export]{adjustbox}

\begin{document}
\title{Shear-Induced Decaying Turbulence in Bose-Einstein Condensates}
\author{S.~Simjanovski}
\affiliation{ARC Centre of Excellence for Engineered Quantum Systems, University of Queensland, Brisbane, AU.}\affiliation{School of Mathematics of Physics, University of Queensland, Brisbane, AU.}

\author{G.~Gauthier}
\affiliation{ARC Centre of Excellence for Engineered Quantum Systems, University of Queensland, Brisbane, AU.}\affiliation{School of Mathematics of Physics, University of Queensland, Brisbane, AU.}

\author{H.~Rubinsztein-Dunlop}
\affiliation{ARC Centre of Excellence for Engineered Quantum Systems, University of Queensland, Brisbane, AU.}\affiliation{School of Mathematics of Physics, University of Queensland, Brisbane, AU.}

\author{M.~T. Reeves}
\affiliation{School of Mathematics of Physics, University of Queensland, Brisbane, AU.}

\author{T.~W.~Neely}
\affiliation{ARC Centre of Excellence for Engineered Quantum Systems, University of Queensland, Brisbane, AU.}\affiliation{School of Mathematics of Physics, University of Queensland, Brisbane, AU.}

\date{\today{}}

\begin{abstract}
    We study the creation and breakdown of a quantized vortex shear layer forming between a stationary Bose-Einstein condensate and a stirred-in persistent current. Once turbulence is established, we characterize the progressive clustering of the vortices, showing that the cluster number follows a power law decay with time, similar to decaying turbulence in other two-dimensional systems. Numerical study of the system demonstrates good agreement of the experimental data with a point vortex model that includes damping and noise. With increasing vortex number in the computational model, we observe a convergence of the power-law exponent to a fixed value.
\end{abstract}

\maketitle
\section{\label{sec:Background}Introduction}

In two-dimensional (2D) turbulence, the conservation of vorticity in the Navier-Stokes equations of motion prompted the use of equilibrium statistical mechanics to predict the end-state configurations of the system. Initiated by Onsager's original ideas more than 70 years ago~\cite{onsager1949statistical}, \and further developed by others~\cite{gauthier2019giant,johnstone2019evolution,sachkou2019coherent}, this approach has resulted in an extensive body of research based on the dynamics of discrete vortices. Recently, experimental results in quantum fluids closely matched the predictions of statistical theories of point vortices~\cite{Reeves_2022}.  

While the statistical approach is quite powerful, for classical 2D fluids the equilibrium predictions are not usually achieved. The observations are that the system remains trapped in quasi-equilibrium states different from the maximum entropy theory predictions~\cite{Tabeling_2002,kurien2001new}, although these configurations may maximize entropy within regions of the system~\cite{jin1998regional}. These quasi-equilibrium states are often vortex crystal states, consisting of separate, ordered regions of intense vorticity in a diffuse vortex background~\cite{jin1998regional,jin2000characteristics,adriani2018clusters,Schecter_1999}. Vortex crystal states are prominent in systems such as planetary atmospheres~\cite{adriani2018clusters} and magnetically guided plasmas~\cite{driscoll2002vortex,Schecter_1999,Schecter_1999}. The initial state for the plasma experiments is a ring of charge, a so-called `hollow-vortex', given the mapping between plasma models and inviscid Euler flow~\cite{driscoll1990experiments}. Subsequent dynamics rapidly rearrange the particles, resulting in the formation of the crystal state. 

For a quantum fluid, the vortices are point-like and a ring of closely-spaced vortices corresponds to a superfluid shear layer. Similarly to a classical fluid, a superfluid shear layer is unstable via an analog to the classical Kelvin-Helmholtz instability (KHI)~\cite{Baggaley_2018,giacomelli2023interplay,HernandezRajkov_2023,Mukherjee_2022}. While observation of the KHI in superfluids has recently been carried out~\cite{Mukherjee_2022,HernandezRajkov_2023}, the end state and dynamics of the ensuing decaying turbulence are still open questions. For classical fluids, decaying 2D turbulence follows a clear trend with power-law decay of the number of concentrated vortices with dependence on time as $N_{C}(t) \propto t^{-\xi}$~\cite{McWilliams_1990,Tabeling_2002}, where the exponent $\xi$ has been observed in the range of $\xi=0.2-1$ depending on the system~\cite{Pomeau_1996,Dezhe_1999,Hansen_1998,Fine_1995,Schecter_1999}.  Theory predicts the end state to be a single cluster, $N_C=1$~\cite{onsager1949statistical, Reeves_2022}. In contrast, classical systems typically exhibit long lived crystal vortex states with $N_C>1$~\cite{Hansen_1998,Schecter_1999}. 
\section{\label{sec:ExperimentalSetup}Experimental Implementation}
\begin{figure}[t!]
    \centering
    \includegraphics[trim={1cm 2.2cm 1cm  2.2cm},clip,width=\hsize]{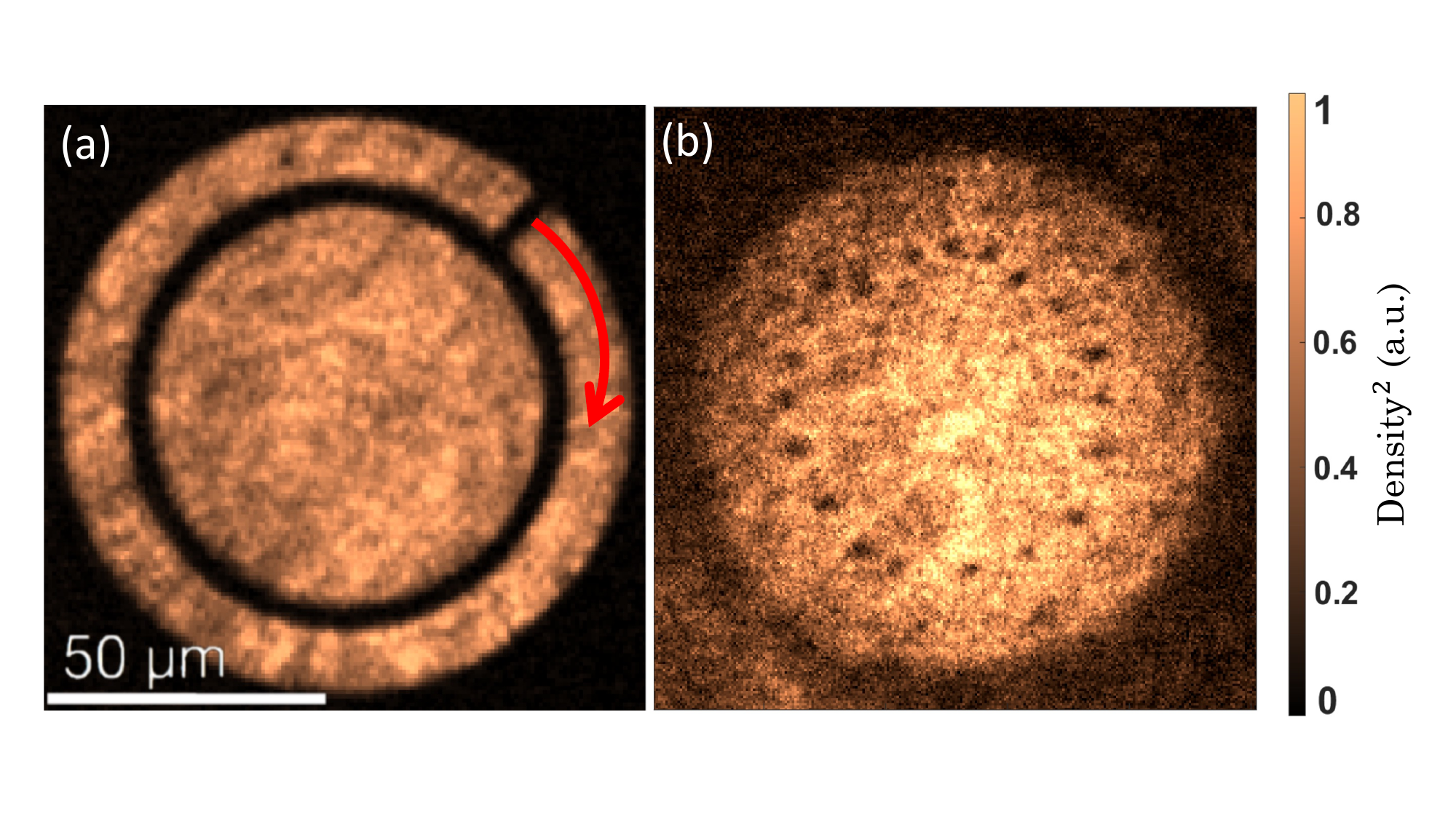}
    \caption{Illustration of the the generation of the shear layer produced using a DMD sequence which initializes the KHI leading to subsequent decaying turbulence. (a) Initial Stirring of the outer region with a separate reference condensate in the center (direction shown by red arrow). (b) Example merging of the outer region and reference condensate by removing the separation barrier creating the shear layer bounded by vortices. Imaged with $5$~ms TOF showing resolved vortices in a ring, here $N=21$ vortices are generated (contrast increased to aid visibility)}.
    \label{fig:Setup}
\end{figure}

Motivated by these points, we experimentally and numerically investigate the dynamics of the decaying turbulence arising from the breakdown of a superfluid shear layer and try to establish whether there exists a similar power law for this situation. The shear layer consists of a ring of twenty quantized vortices and is observed to be unstable in agreement with predictions of the discrete KHI instability~\cite{Aref_1995,HernandezRajkov_2023}. We find that the longer time dynamics are characterized by progressive clustering of the vortices, with a power law dependence of $N_C(t)\propto t^{-0.21\pm0.07}$. At our longer hold times, we find that the vortex system equilibrates to a single on-axis cluster, consistent with our prior work~\cite{Reeves_2022} and maximum entropy theories.

Our experimental system consists of a quasi-2D $^{87}$Rb optically trapped condensate with $N\sim3\times10^6$, where the radial trapping is controlled using high-resolution direct imaging of a digital micromirror device (DMD)~\cite{Gauthier_2016}. In order to observe shear layer decay, we start by initializing shear flow in our system in a ring geometry. We stir in a persistent current in an outer ring region while maintaining a stationary condensate in the center, as shown in Fig.~\ref{fig:Setup}(a)~\cite{HernandezRajkov_2023,Baggaley_2018, Simjanovski_2023}. When the rotating and stationary condensates are recombined, the shear layer is formed, Fig.~\ref{fig:Setup}(b). The stirring process is optimized by utilizing a machine learner to control the stirring parameters, as we describe in a separate publication~\cite{Simjanovski_2023}. Using the parameters of that study, we find a stirring condition which enables consistent generation of a ring comprising of an average of $\overline{N} = 19.6 \pm0.3$ vortices at the shear boundary as shown in Fig.~\ref{fig:Setup}~(b)~\cite{Simjanovski_2023}.

\section{\label{sec:Dynamics}Decay Dynamics}
After initializing the shear layer, we image the condensate after a short $5~$ms time of flight (TOF) for increasing hold times $t_h$, where the system is held within the larger disk trap for the duration $t_h$~\cite{gauthier2019giant,Reeves_2022}. Vortices are detected using a convolutional neural net (CNN) developed by Metz et al.~\cite{Metz_2021}, trained on similar experimental data~\cite{neely2024melting}. In the ensuing dynamics, we observe rapid breakup of the initial vortex ring, followed by  grouping of the vortices into separate clusters. In the following subsections, we focus on analyzing three aspects of the dynamics: (i) the initial decay in the context of the discrete KHI instability, (ii) the clustering dynamics with increasing hold times, and (iii) the long-time equilibration of the system.
\subsection{\label{sec:StructureConstant}Instability dynamics}
\begin{figure}[t!]
    \centering
    \includegraphics[trim={0cm 0cm 0cm  0cm},clip,width=\hsize]{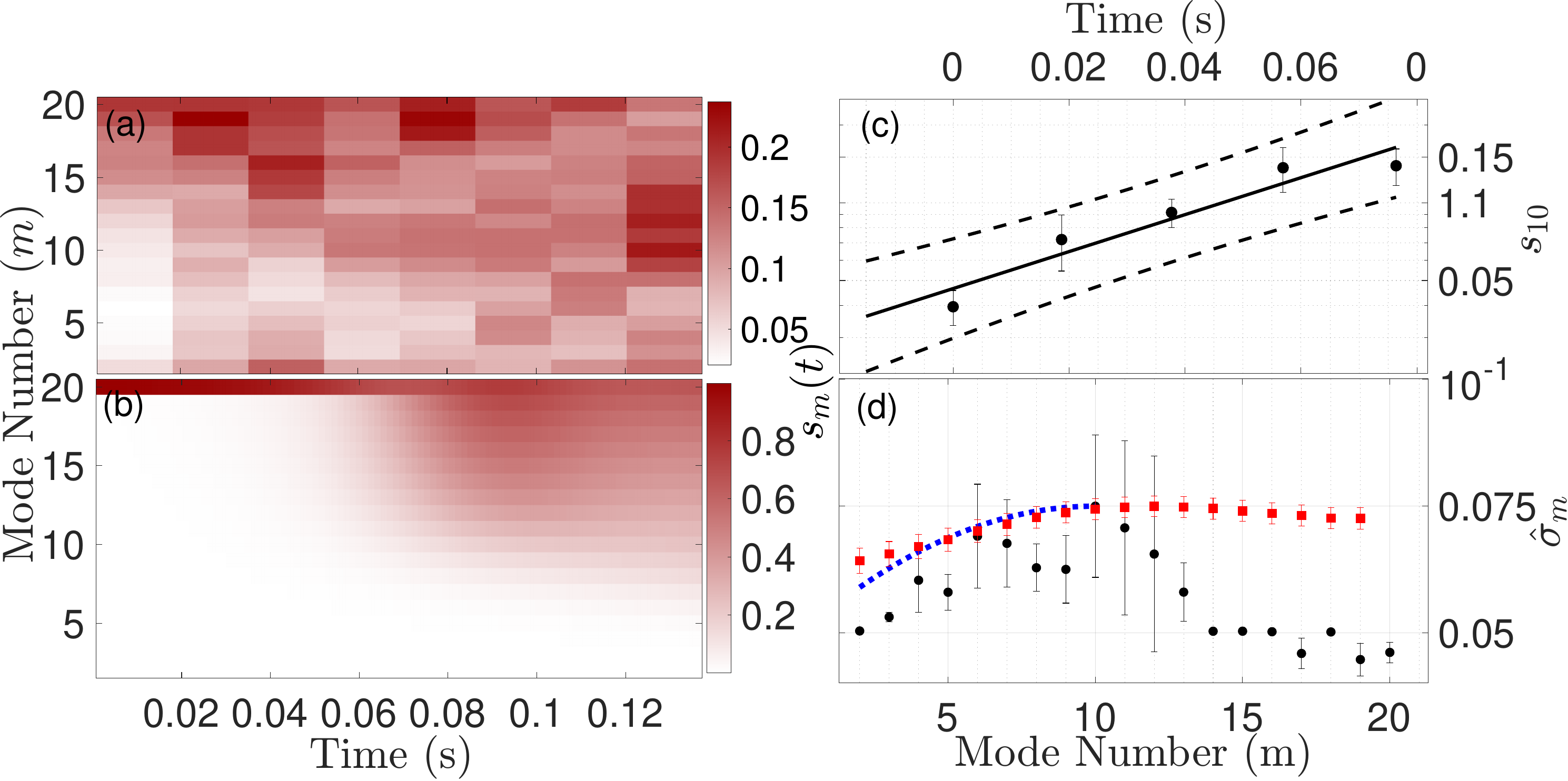}
    \caption{Results of the structure factor analysis for (a) the experimental ring of $N=20$ vortices (the color map is made using an average of all 9 experimental runs) and (b) an ensemble average of 250 SPVM simulations starting with an ideal $N=20$ vortex ring. The shear breakdown is apparent in the growth of modes $m<20$ and occurs later for the ideal case as compared to the experimental initial condition. All results are in the short time domain, spanning over 2-20 modes. (c) Semi-logarithmic plot for experimental mode $m=10$ showing the optimum fitting according to an adaptive fitting algorithm. Square points represent experimental data while the solid line is the line of best fit. Dashed lines show the 95\% confidence interval. Error bars are estimated by the standard error over the nine experimental realizations at each hold time. (d) Normalized growth rates $\hat{\sigma}_m=\sigma_m/\text{max}(\sigma_m)$. Vertical error bars are estimated from the 95\% confidence intervals of the linear fits used to find $\sigma_m$. The dashed (blue) curve shows the predicted growth rates according to a 20 point vortex model following Refs.~\cite{Aref_1995,HernandezRajkov_2023}. The square (red) points show the normalized growth rate according to ideal SPVM ensemble.}
    \label{fig:StructureFactorResults}
\end{figure}

For the case of a shear layer consisting of a ring of discrete vortices, the initial decay is predicted to occur at twice the scale of the inter-vortex spacing~\cite{Aref_1995,Charru_2011,Baggaley_2018}. These initial dynamics are thus dominated by pairing of adjacent vortices. As vortex pairs co-rotate~\cite{Navarro_2013}, the angular distribution of vortices about the center of the initial ring rapidly changes. Hernandez-Rajkov et al.~\cite{HernandezRajkov_2023} recently utilized an angular structure function to characterize these clustering dynamics. The structure function is defined as:

\begin{equation}
    S_m(t)=\frac{1}{N}\sum_{j=1}^{N}\sum_{k\neq j}e^{im[\theta_j(t)-\theta_k(t)]}, \label{eqn:StructureFactor}
\end{equation}

\noindent and is strongly peaked at mode number $m$ when the vortices and/or vortex clusters are azimuthally spaced by $2\pi/m$. 

We apply the analysis of Ref.~\cite{HernandezRajkov_2023} to our data [See Appendix~\ref{sec:StructureFactorFull}]. The normalized structure factor for the experiment, $s_m(t)=S_m(t)/N$, where $N$ is the vortex number, is shown in Fig.~\ref{fig:StructureFactorResults}~(a), focusing on the initial dynamics for $t_h<0.15~$s. As predicted, the mode with the largest amplitude at $t_h=0$~s is $m=20$, corresponding to the $N=20$ separate vortices forming the initial ring structure. As the shear layer breakdown begins, we see this mode decay over time and the other modes are emerging, with the peak of mode growth rate occurring for $m\sim10 \equiv N/2$, in agreement with the results of ef.~\cite{HernandezRajkov_2023}. This is evident in Fig.~\ref{fig:StructureFactorResults}~(d) where we show the normalised growth rates of these modes, denoted by $s_m(t)\propto e^{\sigma_m t}$ (see Appendix~\ref{sec:StructureFactorFull}). The peak growth rate at $m=10$ which suggests that the fastest emerging structure is a collection of vortex pairs since we have $m=N/2$. This indicates that the shear breakdown is dominated by the formation of vortex pairs. We compare these results against theory presented in Ref.~\cite{HernandezRajkov_2023}, which predicts unconditional instability of a row of identical point vortices with uniform spacing $a$, exhibiting decay with growth rates according to~\cite{Aref_1995}:
\begin{equation}
    \sigma_\lambda=\frac{\pi\Gamma}{a\lambda}\left(1-\frac{a}{\lambda}\right), \label{eqn:ArefGrowthRate}
\end{equation}
for a perturbation of wavelength $\lambda$ and fixed vortex circulation $\Gamma$. This model is only valid up to the peak mode number~\cite{Aref_1995}. In order to optimize the fit of this curve, we re-scale the maximum value such as to minimize the error of the resulting curve to the experimental data subject to the shear flow. Doing so gives the dotted (blue) curve in Fig.~\ref{fig:StructureFactorResults}~(d) for which the peak is at $m=10$. 

We compare the experimental data with numerical simulation of a damped stochastic point vortex model (SPVM)~\cite{Reeves_2022,neely2024melting}, where the noise and damping parameters are determined through a best-fit to the experiment, see Appendix~\ref{sec:PVSims}. The SPVM has been previously shown to quantitatively model vortex experiments in the presence of frictional damping and stochastic noise~\cite{Reeves_2022,neely2024melting}. Using a total of 250 SPVM simulations we reconstruct the structure factor as is shown in Fig.~\ref{fig:StructureFactorResults}~(b). Here it is clear that the ring is stable for longer than in the experiment due to the persistence of the $m=20$ mode for the ideal initial conditions of the simulations, consisting of a ring of perfectly spaced vortices. We clearly observe a decaying occupation of the $m=20$ mode for these SPVM results and a gradual occupation of modes $m<20$. Additionally, in Fig.~\ref{fig:StructureFactorResults}~(d) we observe that the SPVM derived growth rates (red squares) follow the theoretical prediction much more closely than our experiment. We attribute these differences to experimental noise which leads to other modes becoming occupied at $t_h=0$, leading to a faster degradation of the ring as well as lowered growth rates since multiple modes both grow and decay simultaneously. 
\subsection{\label{sec:Clustering}Clustering dynamics}
\begin{figure}[t]
    \centering
    \includegraphics[trim={17.5cm 2.1cm 20cm  1cm},clip,width=\hsize]{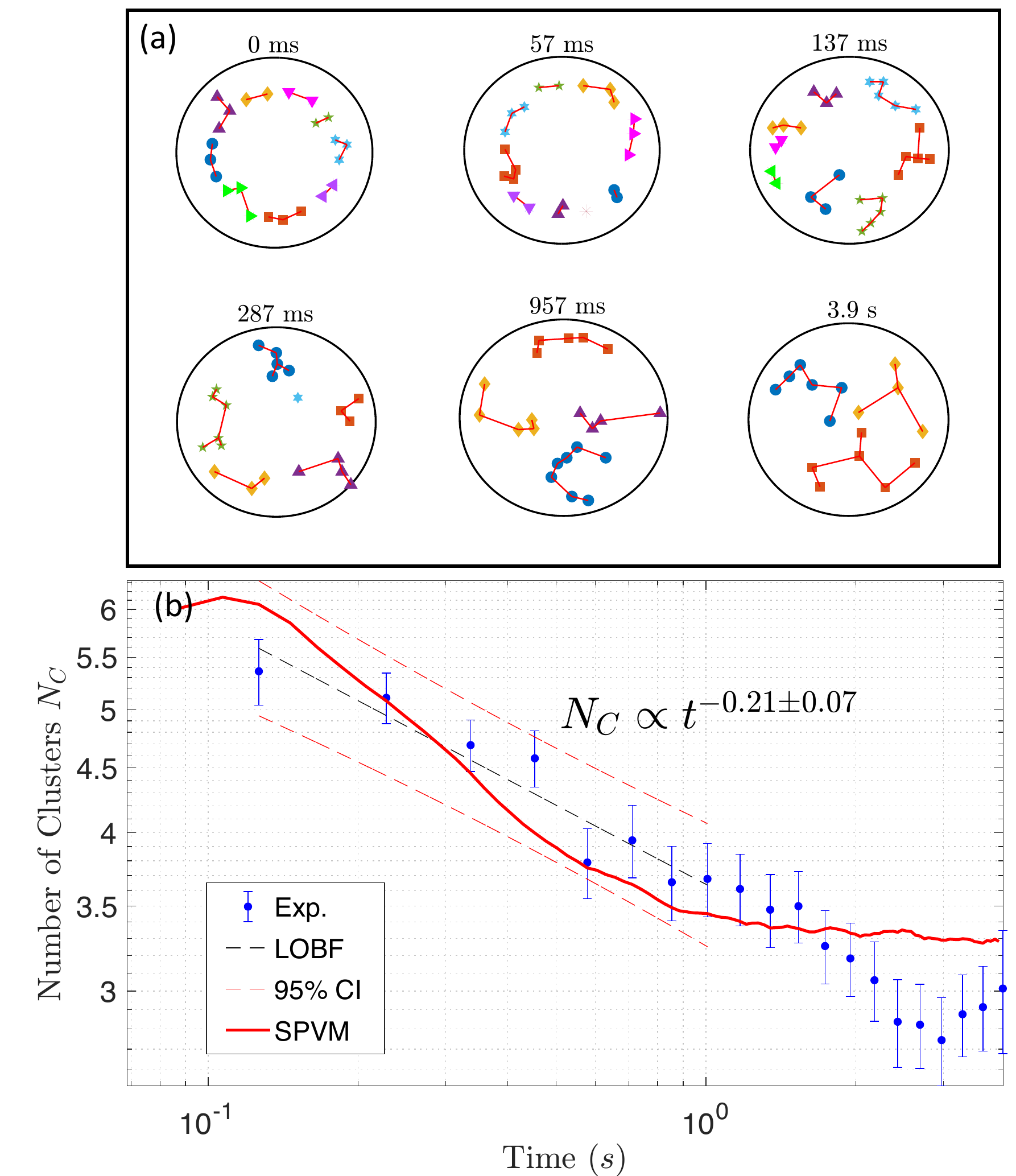}
    \caption{(a) Selected outcomes of clustering using the $\textit{k}$-means algorithm, at the hold times indicated. Points connected via the minimum spanning tree lines belong to the same cluster, also indicated by the identical markers/colors. The black boundary indicates the edge of the BEC. (b) Decay of the vortex cluster number over time according to $\textit{k}$-means algorithm for the experiment (blue dots), and SPVM moving average (solid red line).  Error bars indicate standard error. The dashed line is a line of best fit (LOBF) indicating power-law decay for the experimental points, with dash-dotted lines indicating the 95\% confidence interval (CI) for the fit. In the experiment the clustering began at earlier times than the SPVM results so the SPVM curve was shifted, ensuring that the peak of the point vortex curve and experimental results occur at the same time.}
    \label{fig:Clustering}
\end{figure}
Following the initial pairing of vortices, it is predicted that the breakdown of the shear layer is characterized by organization of the vortices into clusters with increasing hold time~\cite{Baggaley_2018}. We use a cosine-metric $\textit{k}$-means algorithm~\cite{Romesburg_1984} as an automated method for analyzing the vortex position data for evidence of clustering. The algorithm is discussed in detail in Appendix~\ref{sec:Silhouette Score}. Qualitatively, the result of the $\textit{k}$-means analysis is progressive clustering with time, as seen in Fig.~\ref{fig:Clustering}~(a), where clusters can be seen to decrease in number but increase in size, i.e. more vortices per cluster at longer hold times.  The progressive clustering resulting from the decay of the shear layer shows similarities to the roll-up of the vortex sheet under the KHI where the length-scale of the vorticity patches increases over time~\cite{Glosli_2007}. 

Figure~\ref{fig:Clustering}~(b) shows the mean clustering results from the nine sets of experimental data and an ensemble average of 3600 SPVM simulations. The SPVM data is further processed with a moving average, with an averaging window of $\sim50$~ms. For both the experimental and SPVM results, a decay of the cluster number $N_C$ is seen, consistent with progressive clustering, where smaller clusters merge over time into larger clusters. The SPVM results agree well with the experiment, remaining within the 95\% confidence interval of the experimental data. Both cases are shown to plateau towards a constant value of $N_C$ near the end of the experimental hold times. 

We find that the cluster number decay is consistent with power-law behavior. For the experimental results, this power law is found to be $N_C\propto t^{-0.21\pm0.07}$ as is shown on the line of best fit. Similar analysis of the SPVM determines a power law fit  $N_C\propto t^{-0.25\pm0.04}$, consistent with the experimental value. We note that the linear fitting used to extract these values is truncated at $\sim 1$~s where we observe the plateau of the cluster number. In the following section, we investigate the stagnation of the cluster number, by comparing the states of the system at these times with that predicted by statistical equilibrium.

\begin{figure}[t]
    \centering
    \includegraphics[trim={0cm 0cm 0cm  0cm},clip,width=\hsize]{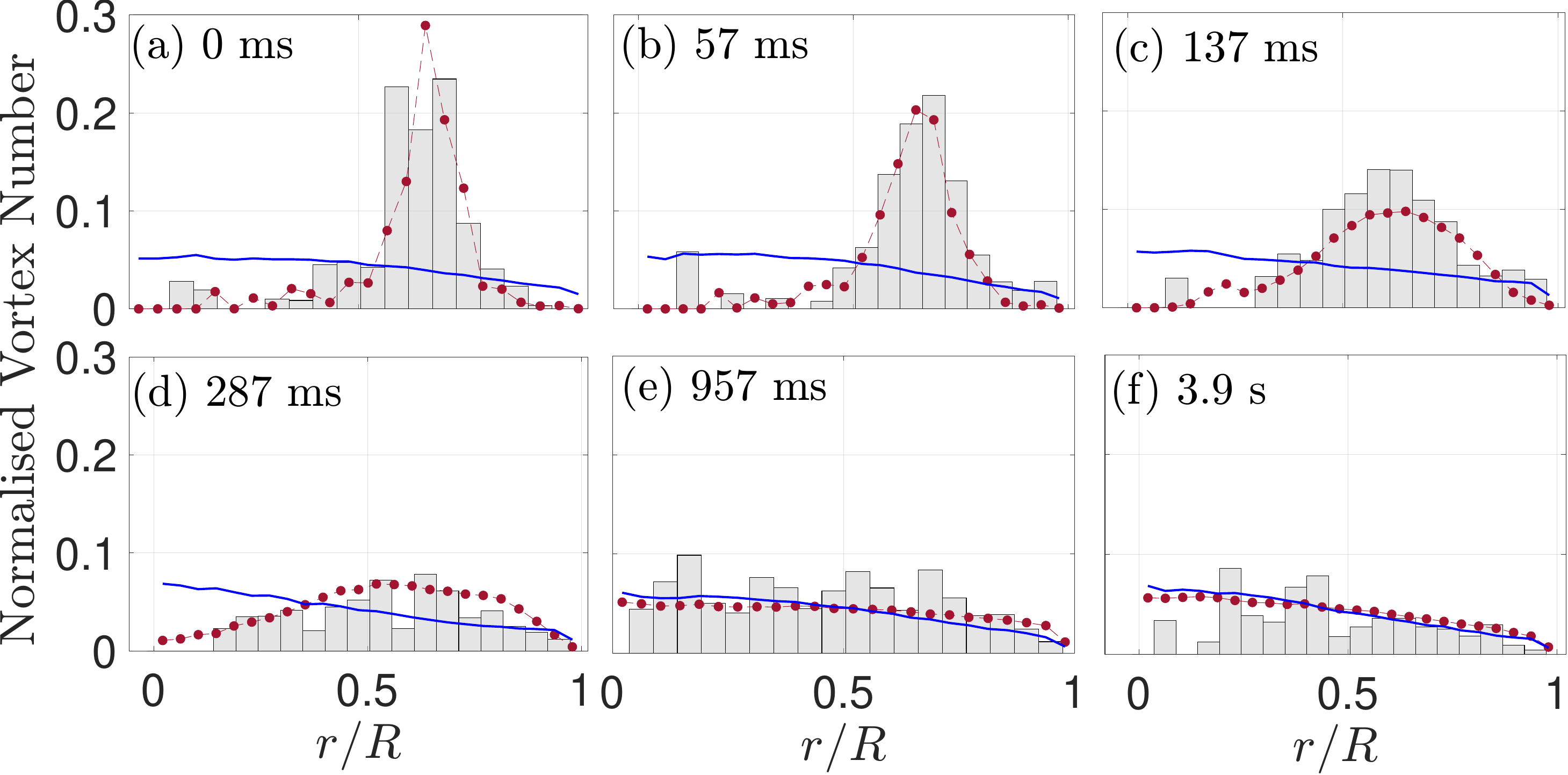}
    \caption{Radial histograms of vortex positions over the normalised disk radius $r/R$. Histograms correspond to experimental times of (a) 0 ms, (b) 57 ms, (c) 137 ms, (d) 287 ms, (e) 957 ms, (f) 3.9 s. The histograms are sampled from: the 9 experimental runs (grey bars); an ensemble of 3600 point vortex simulations, (red dot-dashed line); and the equilibrium estimate (blue solid line). The distributions of (e) and (f) show that the equilibrated state of the vortex system is a single, on-axis cluster [See Appendix~\ref{sec:GaussianFit}].}
    \label{fig:RadialHistogram}
\end{figure}
\subsection{\label{sec:ProlongedEquilibration}Vortex Cluster Equilibration}
The stagnation of the cluster number at $t\gtrsim 1$~s is initially suggestive of the system failing to achieve equilibrium, as a single cluster is the expected end state of the decaying turbulence~\cite{Reeves_2022}. Using the 3600 SPVM realizations of the vortex dynamics, we plot radial histograms of vortex positions in Fig.~\ref{fig:RadialHistogram} for the numerical results (red dot-dashed line) and experiment (grey bars). These are compared with estimates of the equilibrium distribution corresponding to the vortex number, angular momentum, and energy at each time, generated by using conservative time-averaged dynamics of a representative sample of the ensemble. Since the point-vortex model is ergodic for small vortex number~\cite{Robert_1991,Dritschel_2015}, the time evolution will explore the available phase-space, estimating the equilibrium distribution~\footnote{Prior work, Ref.~\cite{Dritschel_2015}, suggests that the point vortex model is ergodic for moderate N. However, other work such as Ref.~\cite{Weiss_1991} demonstrate the non-ergodicity of large N point vortex systems.}.

As shown in Fig.~\ref{fig:RadialHistogram}, the SPVM dynamics closely matches the equilibrium estimate by $t\sim957$~ms, suggesting that vortex system has equilibrated well before the end of the experimental run time. We thus conclude that the stagnation in cluster number after $t \gtrsim 1$~s indicates a breakdown in the clustering algorithm's efficacy at later times. In other words, we expect that the distributions in Fig.~\ref{fig:Clustering}(a) for $t \gtrsim 1$~s correspond to single clusters (on average). We thus choose a conservative cut-off for our power law fiting in Section~\ref{sec:Clustering} at $1$~s. 
\section{\label{sec:Conclusion}Discussion and Conclusion}
In summary, we find that shear layer breakdown results in the initial formation of vortex pairs, in agreement with prior work and theory~\cite{HernandezRajkov_2023,Robert_1991}. The subsequent decaying turbulence at longer times is characterized by the progressive clustering of the vortices, with the cluster number exhibiting power law decay $N_C(t)\propto t^{-\xi}$. Our result for the exponent, $\xi=0.21\pm0.07$, falls within the classically known range for KHI driven decay rates, $\xi=0.2-1$~\cite{Pomeau_1996,Dezhe_1999,Hansen_1998,Fine_1995,Schecter_1999}. Despite the power-law decay, we find no evidence of delayed equilibration in our experiment. This is in contrast to larger classical plasma systems~\cite{Schecter_1999}, and may be due to the relatively small system size and small vortex number, or perhaps due to the dominance of the stochastic noise in driving the system dynamics.

A closely analogous system to discrete superfluid vortices is the 2D Euler flow of electrons~\cite{Schecter_1999}, where the number of concentrated vortices was seen to decay with $N_C \propto t^{-0.25}$, although we note that our approach of defining vortex clusters via k-means differs from Ref.~\cite{Schecter_1999}. However, since the electron experiments contain $N>10^{5}$ particles, we used our numerical model to investigate the dependence of our power law exponent for superfluid vortex clusters on initial vortex numbers. In Appendix~\ref{sec:LargerNumberPVSimulations}, additional details are provided on the dynamics of $N=50$, $N=100$ and $N=200$ ensembles, starting from the same initial vortex ring diameter as the $N=20$ case. We find that the decay exponent is $\xi=0.6\pm0.1$, $\xi=0.68\pm0.06$ and $\xi=0.67\pm0.03$ for $N=50$, $N=100$ and $N=200$ vortices respectively, indicative of convergence towards a fixed value of $\xi$ as N is increased. As described in the Appendix D, the most unstable mode in all cases is again $m \sim N/2$, consistent with expectations from the point vortex theory of the KHI~\cite{Aref_1995,HernandezRajkov_2023}. 

Future directions for this work may involve increasing the vortex number in the experiment, while also suppressing the noise term which significantly affects the vortex dynamics~\cite{Reeves_2013}. Prior numerical work predicts a delayed equilibration for an initial point vortex ring in the absence of noise and limited damping~\cite{Stockdale_2020}, qualitatively consistent with the vortex crystal states observed in the electron systems~\cite{Schecter_1999}. Observing these non-equilibrium states and delayed equilibration experimentally are a clear future directions for this research. The high level of control over the experimental system suggests future work in investigating the dynamics of the KHI itself, such as probing the influence of the boundary conditions on the instability rate, as the proximity of the vortex ring to the system boundary is predicted to change the instability dynamics~\cite{Giacomelli_2023,caldara2024suppression}. A double ring geometry, including a central pinning site~\cite{Baggaley_2018,HernandezRajkov_2023}, can also introduce effects due to a non-zero pinned circulation in the ring. This is predicted to play an important role in the instability dynamics, potentially stabilizing the breakdown for particular vortex numbers~\cite{havelock1931lii, HernandezRajkov_2023}. 
\section{Acknowledgments}
We thank Matthew J. Davis for useful discussions. This research was supported by the Australian Research Council (ARC) Centre of Excellence for Engineered Quantum Systems (EQUS, CE170100009). S.~S~acknowledges the support of an Australian Government Research and Training Program Scholarship. G.~G.~acknowledges the support of ARC Discovery Projects Grant No. DP200102239. M.T.R is supported by an Australian Research Council Discovery Early Career Researcher Award (DECRA), Project No. DE220101548. T.~W.~N.~acknowledges the support of Australian Research Council Future Fellowship No. FT190100306.
\appendix
\section{\label{sec:StructureFactorFull}Detailed Structure Factor Analysis}
In order to quantitatively extract the mode growth rates from the structure factor, we perform linear regression on a semi-logarithmic scale for each mode number. This fitting must be performed carefully, since only for short times will the structure factor show rapidly growing modes. Performing a fit on the entire time series will result in a relatively uniform distribution of $\sigma_m$ near zero. However, the time series that was collected in the experiment, being focused on later clustering dynamics, does not provide enough resolution at early times to apply a standard fitting protocol to extract the growth rate. Instead we apply an adaptive fitting algorithm, where the $R^2$ values for a series of linear fits, iterating the final fitting time towards the maximum experiment time. The maximum $R^2$ value for each fit and mode number is used to find the most appropriate time to truncate the fitting.

For slow growing modes such as $m=1-5$, the growth is approximately flat over the entire time as shown in Fig.~\ref{fig:StructureFactorResults}~(b), meaning a growth rate near zero. On the other hand, fast growing modes are truncated after very short times as their maximum values are quickly reached. This results in larger uncertainties for the growth rates associated with the fast growing modes. The uncertainties are estimated via the 95\% confidence interval of the slopes extracted from the linear fitting. The results are shown in Fig.~\ref{fig:StructureFactorResults}~(b) where the peak of the growth rate distribution is seen near $m=10$~consistent with theory predictions~\cite{Aref_1995}.
\section{\label{sec:PVSims}Point Vortex Simulations}
\begin{figure}[t]
    \centering
    \includegraphics[width=\hsize]{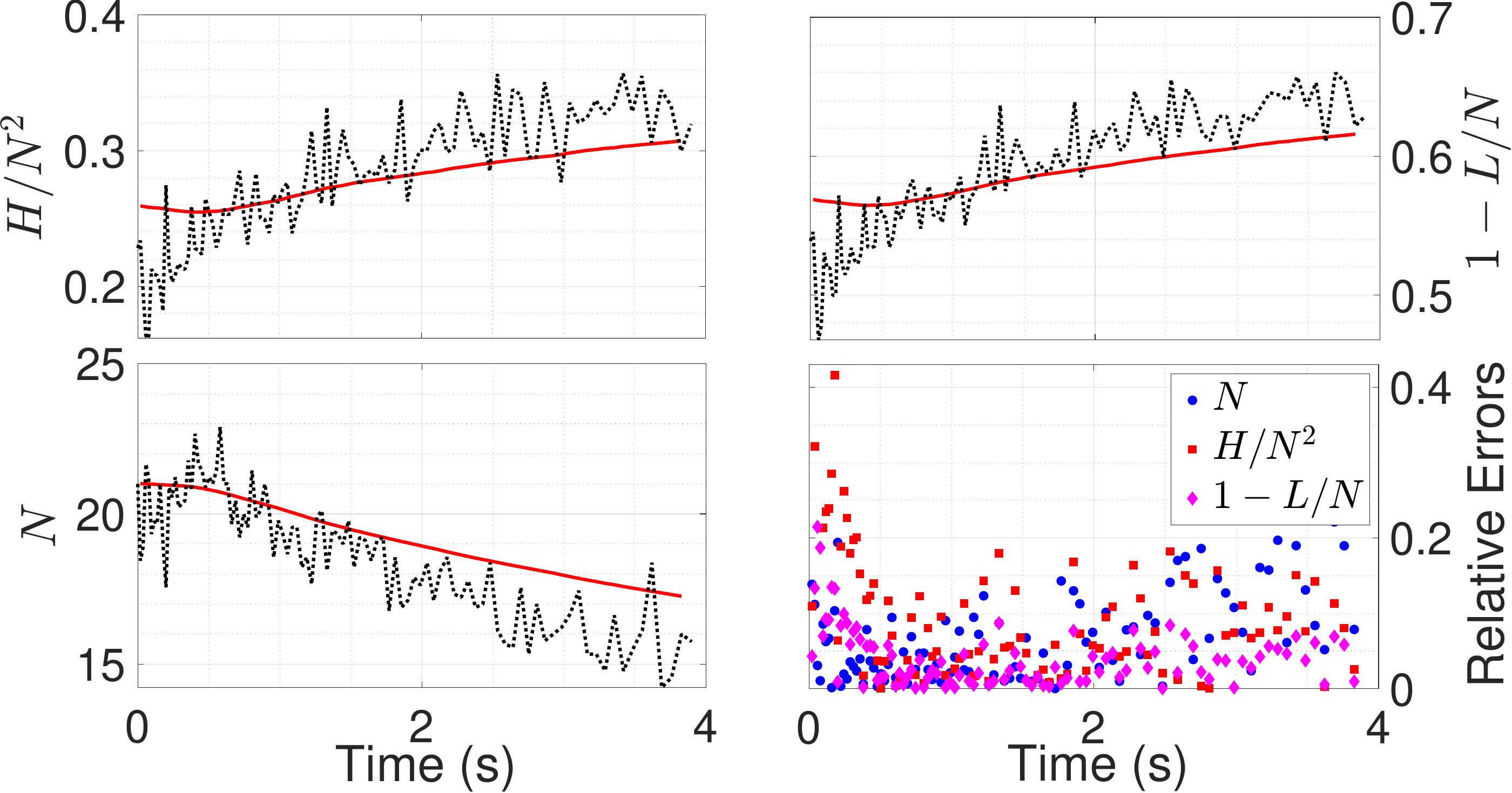}
    \caption{(a) Vortex number $N$ vs. time for the experimental vortex positions shown by the (black) dotted lines and the SPVM results shown by the (red) solid. (b) Normalised energy, $H/N^2$ vs. hold time. Both experimental and SPVM results are shown similarly to (a). (c) Normalised angular momentum $1-L/N$. (d) The relative error between the energy, angular momentum, and vortex number between the experiment and SPVM results. The optimum parameters for the SPVM modelling were determined by minimizing this error. All plots shown use the optimum point vortex parameters of $\alpha=1.2\times10^{-3}$ and $\eta=10^{-2}$.}
    \label{fig:RelativeErrorPVM}
\end{figure}
We use a stochastic point vortex model (SPVM) to model the dynamics of our system~\cite{Reeves_2022}. Vortex dynamics are parameterized by the equation of motion~\cite{Reeves_2022}:
\begin{equation}
    d\bold{r}_j=[\bold{v}_j-\alpha\hat{\bold{z}}\times\bold{v}_j]dt+\sqrt{2\eta}d\bold{W}_j, \label{eqn:PVMEqMotion}
\end{equation}
where $\bold{r}_j$ and $\bold{v}_j$ represent the position and velocity of the $j$-th vortex respectively, $d\bold{W}_j$ is a complex Gaussian noise variable, $\alpha$ the mutual friction coefficient, and $\eta$ the vortex diffusion rate. The noise strength $\eta$ representing fluctuations in vortex motion induced by fluctuating background sound modes in the superfluid~\cite{Reeves_2022}. Optimal values of $\alpha$ and $\eta$ are determined by generating an error landscape similar to prior work~\cite{Reeves_2022} that compares the values from modelling the dynamics with Eq.~\ref{eqn:PVMEqMotion} with those from the experiment. The error is estimated as a sum of the error in energy per square vortex number $H/N^2$, angular momentum per vortex $1-M/N$ and the vortex number $N$. All these quantities can be found directly from vortex positions, and so the experimental and point vortex values of $H,M,N$ can be compared directly. The optimal parameters that minimize the error between the simulated and experimental $H,M,N$ are found to be  $\alpha=1.2\times10^{-3}$ and $\eta=10^{-2}$.  Individual error contributions under these optimum values is shown in Fig.~\ref{fig:RelativeErrorPVM}(d).
\section{\label{sec:Silhouette Score}Details of the Clustering Algorithm}
Clustering using single shot images requires a restriction of the definition of clustering to ``points close to one another". Ideally, any clustering algorithm would account for vortices in a cluster to be co-rotating which is known from numerical work~\cite{Baggaley_2018}. This is because the co-rotating property of the clusters can be used as an additional selection criteria for characterizing said clusters~\cite{Easton_2023}. Performing this classification solely based on vortex positions is more complicated and can incorrectly characterize a cluster of vortices which are realistically just close to one another at some given instance but do not co-rotate (hence they are not a true cluster). Unfortunately, this is not possible with our destructive imaging, which limits the information accessible in the experiment. Each image is a new instance of the decaying turbulence which, as a chaotic system, cannot be matched directly to prior data to reconstruct the dynamics which would be necessary for determining co-rotation. Thus, a $\textit{k}$-means cluster sorting algorithm is applied to the vortex positions extracted by the CNN detection. This method is combined with a silhouette score~\cite{Rousseauw_1987} to find the cluster number. Our choice of this approach to cluster classification was to apply a relatively unbiased estimate of the cluster number that permits automatic classification of the data without any fine-tuning of input parameters.
\section{\label{sec:LargerNumberPVSimulations}Larger Number Point Vortex Simulations}
 \begin{figure}[t!]
    \centering
    \includegraphics[trim={23cm 0cm 23cm  0cm},clip,width=\hsize]{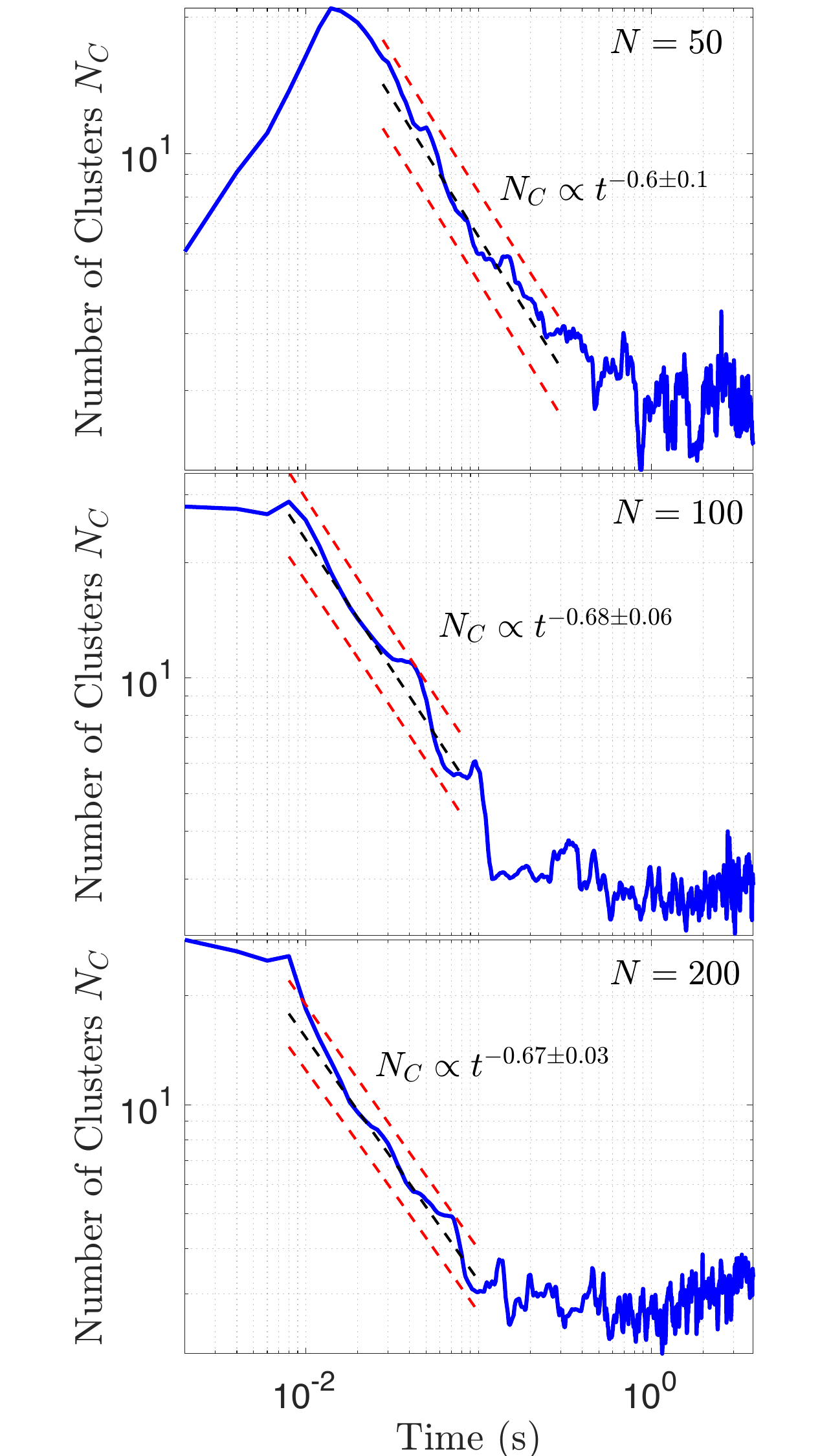}
    \caption{Power law decay of the cluster number over time observed through the use of $\textit{k}$-means algorithm combined with the silhouette score (Section~\ref{sec:Clustering}) in the case of (a) $N=50$ and (b) $N=100$ vortices in a ring. The solid (red) curves indicate the moving average of the cluster number $N_C$ for the first $136.$~ms before equilibration has occurred. The dotted (blue) curve shows the moving average for the cluster number after equilibration, where the clustering algorithm is ineffective. The dashed (black) line shows the line of best fit and the dashed (orange) lines show the $95\%$ confidence bounds are shown for both the point vortex simulations and the line of best fit. The fit is restricted to early times due to the rapid equilibration in these large vortex cases.}
    \label{fig:LargeNVortexDecay}
\end{figure}
 \begin{figure}[t!]
    \centering
    \includegraphics[trim={0cm 0cm 0cm  0cm},clip,width=\hsize]{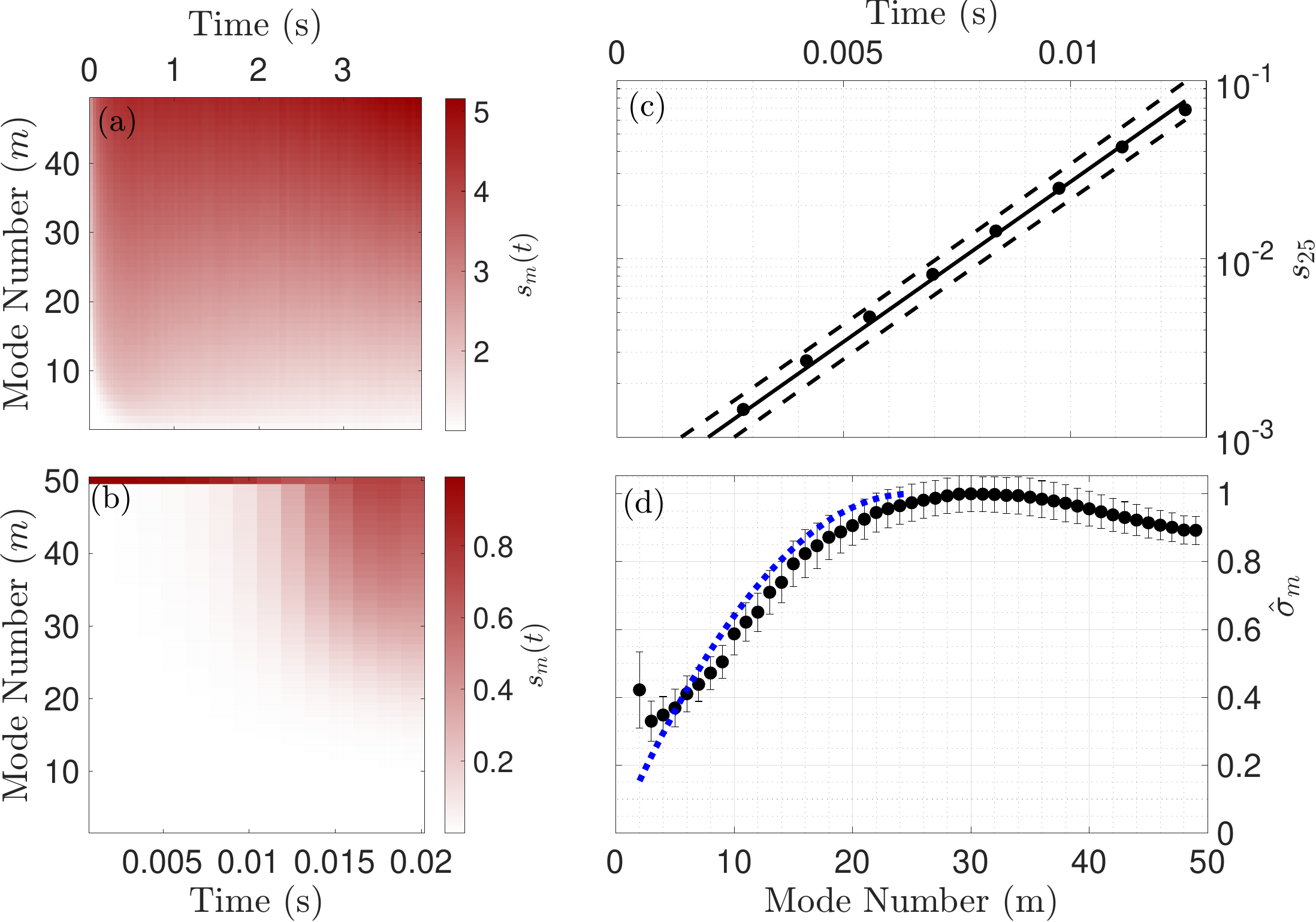}
    \caption{Structure factor analyis for $N=50$ vortices simulated by the SPVM in an ideal ring. (a) Complete structure factor over $3.9$~s (matching experimental times). (b) Structure factor over $20$~ms, used for finding initial fast growth rates. (c) Fitted structure factor over time for mode $m=25=N/2$, where the slope indicates growth rate. (d) Growth rates for each mode number, points (black). A theory curve consistent with Eq.~\ref{eqn:ArefGrowthRate} is shown by the dotted (blue) curve.}
    \label{fig:LargeNSTF}
\end{figure}
The algorithm works by partitioning the data set of vortex positions, into $k$ subsets such that the variance of each set from its mean is minimized~\cite{MacQueen_1967}. Typically, this requires $k$, the cluster number, to be specified, which is not known here beforehand. Instead, we sweep through a range of $k$ values and assign each sorted data point, $i$, a silhouette measure, $s_k(i)$, based on its similarity to other points both within its cluster, denoted by $a(i)$ and outside of its cluster, $b(i)$~\cite{Rousseauw_1987}. Typically, $a(i),b(i)$ are just square Euclidean distances between the points, although we employ an alternative method discussed further below. The silhouette is defined as follows:
\begin{equation}
    s_k(i)=\frac{a(i)-b(i)}{\text{max}\{a(i),b(i)\}}\label{eqn:silhouette}.
\end{equation}
Note that Eq.~\ref{eqn:silhouette} ensures that $-1\leq s_k(i)\leq1$, where the closer the silhouette is to $1$, the more certain we can be that the cluster contains the data point $i$. $s_k(i)\to-1$ denotes the opposite of this~\cite{Rousseauw_1987}. From $s_k(i)$, we can also assign a score $S_k=\text{mean}\{s_k(i)\}$ as an average measure of the performance of the $\textit{k}$-means algorithm for the given $k$ choice. Repeating this process for various $k$, we find the best fit to be that which maximizes $S_k$. Our choice of similarity score is a variation on the cosine metric~\cite{Romesburg_1984}. We consider this measurement, $d$, to be related to the cosine of the included angle between some vortex position, $\vec{r}$, relative to a cluster centroid $\vec{c}$ via:
\begin{equation}
    d(r,c)=1-\frac{\vec{r}\cdot\vec{c}}{\sqrt{|\vec{r}|^2|\vec{c}|^2}}. \label{eqn:CosineMetric}
\end{equation}
The centroid positions are initially estimated and iteratively updated with each clustering event using the so-called ``$\textit{k}$-means++ algorithm"~\cite{Arthur_2007}. The cosine metric is chosen due to the initial ring structure of the vortex shear, meaning that the breakdown of the shear layer is most notable through angular displacements of the vortex positions. This does however limit the accuracy of the clustering algorithm at long hold times, where radial displacements are important to consider. Examples at various hold times are shown in Fig.~\ref{fig:Clustering}~(a) with connected points representing clustered vortices. We note the limitation in the algorithm is also highlighted here for the example classification at $3.9$~s, where the cosine metric prevents the three vortices near the center of the system to be classified as a cluster due to their larger azimuthal displacements. Aside from these extremely late times, the algorithm functions well enough to give a detailed view of the cluster behavior resulting from the shear layer breakdown as is indicated by the other examples in Fig.~\ref{fig:Clustering}~(a).

For point vortex simulations requiring vortex numbers different than those considered in the experiment, an initial ring of equally spaced vorticies is used. Following this, the position of each vortex in the ring is shifted randomly by up to $1$\% of it radial and angular position. This is sufficient to generate the decaying shear layer discussed in Section~\ref{sec:Clustering}. Specifically, starting with $N=50$ vortices in the ring gives a clear power law decay $N_C\propto t^{-0.6\pm0.1}$ as shown in Fig.~\ref{fig:LargeNVortexDecay}. In the case of $N=100$ vortices we find that $N_C\propto t^{-0.68\pm0.06}$. Finally for $N=200$ we find that $N_C\propto t^{-0.67\pm0.03}$. These results suggest that the power law decay of the progressive clustering from the decaying turbulence converges to a fixed value for a large enough system, i.e. as the continuum limit is approached. 

For completeness, we also repeat the structure factor analysis, Eq.~\ref{eqn:StructureFactor}, for the case of $N=50$ vortices. This is shown in Fig.~\ref{fig:LargeNSTF}, where we see similar behaviour to the ideal $N=20$ case shown in Fig.~\ref{fig:StructureFactorResults}~(b). The decay of the $m=50$ into lower mode numbers at short times has a short period of stability initially, although clearly the breakdown of the shear begins sooner in this case, within $20$~ms. We see decent agreement of the subsequent growth rates for each mode to the theoretical predictions of Eq.~\ref{eqn:ArefGrowthRate}, with the peak mode growth being near $m=N/2=25$, within error due to the fitting algorithm. 
\section{\label{sec:GaussianFit}Fitting the Equilibrium State}
\begin{figure}[t!]
    \centering
    \includegraphics[width=\hsize]{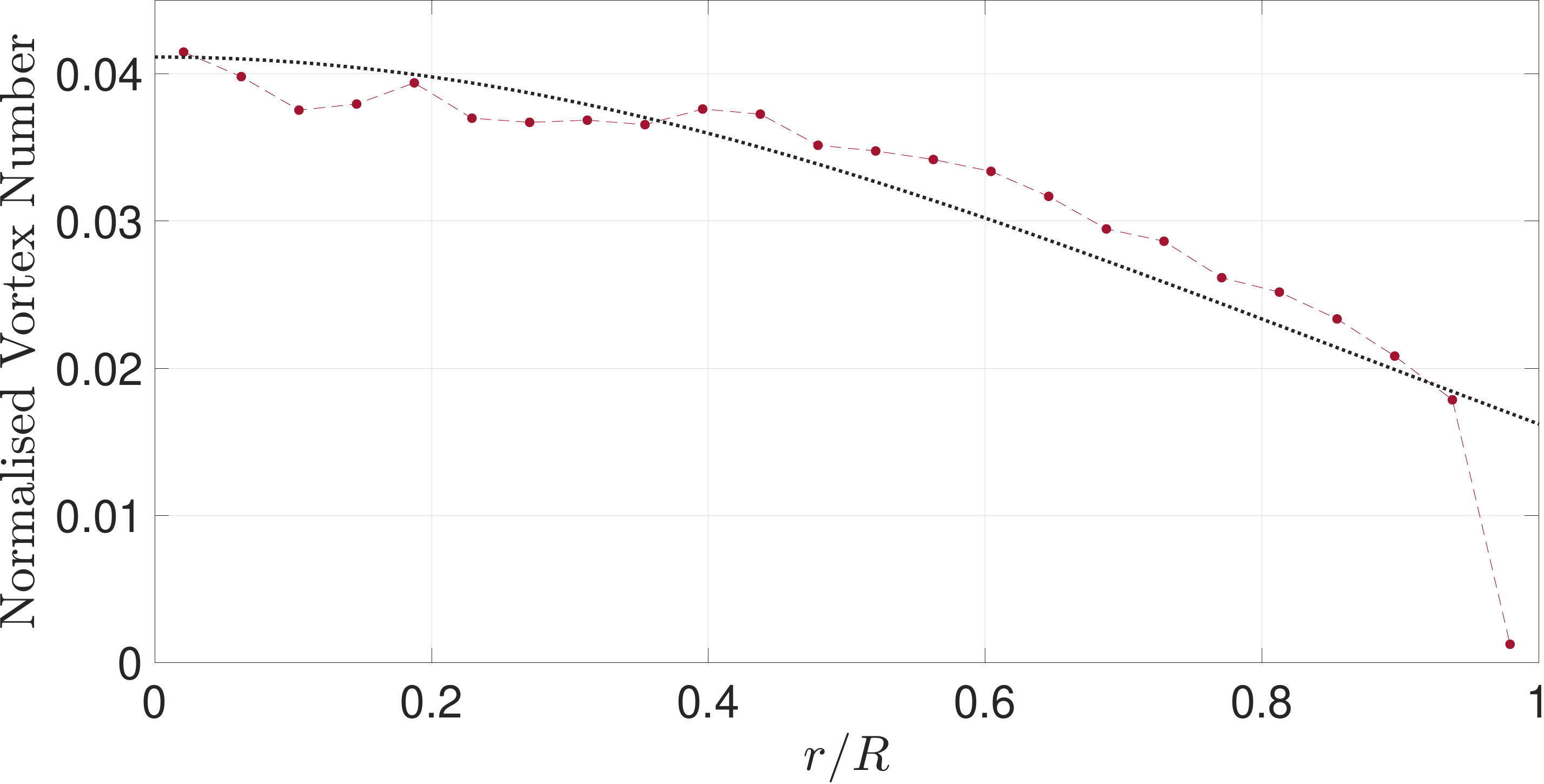}
    \caption{A Gaussian fit, shown by the dotted black curve, to the SPVM ensemble (red-dashed) at $0.957$~ms.}
    \label{fig:GaussianFitSPVM}
\end{figure}

If the system achieves equilibrium, we expect the state to be an on-axis cluster~\cite{Reeves_2013,Tornkvist_1997,smith_1990}, which has approximately a Gaussian character in the positive vortex temperature regime. In Fig.~\ref{fig:GaussianFitSPVM} shows a Gaussian fit centered at the origin to the ensemble average of the SPVM simulations at $0.957$~ms. We observe this is consistent with a vortex distribution for a single, on-axis cluster, supporting our conclusion that the system has reached equilibrium at this stage of the evolution.
\bibliography{references}
\end{document}